\documentclass{elsart}
\usepackage{natbib}
\usepackage[dvips]{graphicx,color}
\usepackage{amsmath,bm}
\DeclareGraphicsExtensions{.eps}
\bibpunct{[}{]}{,}{n}{}{;}
\def\bT{{\bm T}}
\def\bL{{\bm L}}
\def\cA{\mathcal{A}}\def\cC{\mathcal{C}}
\def\cR{\mathcal{R}}\def\cZ{\mathcal{Z}}
\def\tm{\tilde{m}}\def\tk{\tilde{k}}
\def\rF{{\rm F}}
\def\pymod#1{~({\rm mod}\,#1)\ }
\def\ket#1{\mid~\!\!\!{#1}~\!\!\rangle} \def\bra#1{\langle~\!\!{#1}~\!\!\!\mid}
\def\emline#1#2#3#4#5#6{%
       \put(#1,#2){\special{em:moveto}}%
       \put(#4,#5){\special{em:lineto}}}
\def\newpic#1{}
\begin{document}
\runauthor{Damnjanovi\'c, Vukovi\'c, Milo\v sevi\'c}
\begin{frontmatter}
 \title{Fermi level quantum numbers and secondary gap of conducting carbon nanotubes}
 \author{M. Damnjanovi\'c\thanksref{emy}}
 \author{T. Vukovi\'c}
 \author{I. Milo\v sevi\'c}
 \address{Faculty of Physics,
 University of Belgrade, POB 368, Belgrade 11001, Yugoslavia,
 http://www.ff.bg.ac.yu/qmf/qsg\_e.htm}
\thanks[emy]{E-mail: yqoq@afrodita.rcub.bg.ac.yu}
\date{\today}
 \begin{abstract}
For the single-wall carbon nanotubes conducting in the simplest tight
binding model, the complete set of line group symmetry based quantum
numbers for the bands crossing at Fermi level are given. Besides
linear ($k$), helical ($\tilde{k}$ and angular momenta, emerging from
roto-translational symmetries, the parities of $U$ axis and (in the
zig-zag and armchair cases only) mirror planes appear in the
assignation. The helical and angular momentum quantum numbers of the
crossing bands never vanishes, what supports proposed chirality of
currents. Except for the armchair tubes, the crossing bands have the
same quantum numbers and, according to the non-crossing rule, a
secondary gap arises, as it is shown by the accurate tight-binding
calculation. In the armchair case the different vertical mirror
parity of the crossing bands provides substantial conductivity,
though $k_\rF$ is slightly decreased.
 \end{abstract}
 \end{frontmatter}
Since their discovery \cite{IIJIMA} conducting properties of carbon
nanotubes have been extensively studied, due to expected
technological applications \cite{DRESSEL-IIJ}. The electronic bands
have been found within the tight binding approximation and assigned
either by the linear $k$ and the angular $m$ quasi momenta quantum
numbers \cite{HAM92,DRE98b}, or by the quantum numbers $\tk$ and
$\tm$ of the helical and "pure" (i.e. the remnant being not coupled
to translations) angular momentum \cite{WHI93,JIS95}. The both sets
of quantum numbers reflect translational and rotational tube
symmetries \cite{YITR}, forming the group $\bL^{(1)}$. The local
curvature has been partly neglected in these calculations, assuming
the same interaction of carbon atom with its three neighbors. This
model predicts \cite{HAM92,WHI93,KAN97} that tube $(n_1,n_2)$ is
conducting whenever $n_1-n_2$ is multiple of 3, what is well
experimentally justified \cite{EXP}. It is also suggested
\cite{HAM92,WHI93} that even in this case the local curvature opens a
small secondary gap (less then 0.01\,eV and decreasing with the
square of the tube diameter \cite{KAN97}) when $n_1-n_2$ is
non-vanishing; this feature seems to be still beyond the experimental
precision.

Here, we present the full set of symmetry quantum numbers of the
bands crossing at the Fermi level within the mentioned simple model.
They are used to show that the symmetry gives the most profound and
easy prediction of the opening of the secondary gap: non-crossing
rule prevents conductivity in all but the armchair tubes. The
conductivity of the later is due to the extra parity contained in the
recently determined \cite{YITR} full symmetry group, which makes
non-crossing rule inapplicable. Finally, the secondary gap is
estimated within a distortion sensitive tight-binding model.

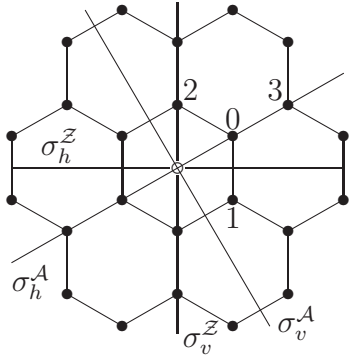
\begin{figure}\centering
\unitlength=0.70mm \special{em:linewidth 0.4pt} \linethickness{0.4pt}
\begin{picture}(68.50,65.00)
\put(4.50,28.00){\circle*{2.00}} \put(4.50,40.00){\circle*{2.00}}
\
\put(15.00,10.00){\circle*{2.00}} \put(15.00,22.00){\circle*{2.00}}
\put(15.00,46.00){\circle*{2.00}} \put(15.00,58.00){\circle*{2.00}}
\
\put(25.50,4.00){\circle*{2.00}} \put(25.50,28.00){\circle*{2.00}}
\put(25.50,40.00){\circle*{2.00}} \put(25.50,64.00){\circle*{2.00}}
\
\put(36.00,10.00){\circle*{2.00}} \put(36.00,22.00){\circle*{2.00}}
\put(36.00,46.00){\circle*{2.00}} \put(36.00,58.00){\circle*{2.00}}
\
\put(46.50,4.00){\circle*{2.00}} \put(46.50,28.00){\circle*{2.00}}
\put(46.50,40.00){\circle*{2.00}} \put(46.50,64.00){\circle*{2.00}}
\
\put(57.00,10.00){\circle*{2.00}} \put(57.00,22.00){\circle*{2.00}}
\put(57.00,46.00){\circle*{2.00}} \put(57.00,58.00){\circle*{2.00}}
\
\put(67.50,28.00){\circle*{2.00}} \put(67.50,40.00){\circle*{2.00}}
\
\emline{4.50}{40.00}{41}{15.00}{46.00}{42}
\
\emline{4.50}{28.00}{43}{15.00}{22.00}{44}
\
\emline{15.00}{58.00}{3}{25.50}{64.00}{4}
\
\emline{15.00}{10.00}{5}{25.50}{4.00}{6}
\emline{15.00}{46.00}{7}{25.50}{40.00}{8}
\
\emline{25.50}{4.00}{9}{36.00}{10.00}{10}
\emline{25.50}{40.00}{11}{36.00}{46.00}{12}
\
\emline{25.50}{28.00}{13}{36.00}{22.00}{14}
\emline{25.50}{64.00}{15}{36.00}{58.00}{16}
\
\emline{36.00}{22.00}{17}{46.50}{28.00}{18}
\emline{36.00}{58.00}{19}{46.50}{64.00}{20}
\
\emline{36.00}{10.00}{21}{46.50}{4.00}{22}
\emline{36.00}{46.00}{23}{46.50}{40.00}{24}
\
\emline{46.50}{4.00}{25}{57.00}{10.00}{26}
\
\emline{46.50}{28.00}{29}{57.00}{22.00}{30}
\emline{46.50}{64.00}{31}{57.00}{58.00}{32}
\
\emline{57.00}{22.00}{33}{67.50}{28.00}{34}
\
\emline{57.00}{46.00}{35}{67.50}{40.00}{36}
\
\put(4.50,28.00){\line(0,1){12.00}}
\
\put(15.00,10.00){\line(0,1){12.00}}
\put(15.00,46.00){\line(0,1){12.00}}
\
\put(25.50,28.00){\line(0,1){12.00}}
\
\put(36.00,10.00){\line(0,1){12.00}}
\put(36.00,46.00){\line(0,1){12.00}}
\
\put(46.50,28.00){\line(0,1){12.00}}
\
\put(57.00,10.00){\line(0,1){12.00}}
\put(57.00,46.00){\line(0,1){12.00}}
\
\put(67.50,28.00){\line(0,1){12.00}}
\
 \put(36.00,34.00){\circle{2.00}}
 \put(4.50,34.00){\line(1,0){30}}
 \put(37.50,34.00){\line(1,0){30}}
 \put(36.0,35.50){\line(0,1){30}}
 \put(36.0,32.50){\line(0,-1){30}}
 \emline{4.50}{16}{1}{67.5}{52}{2}
 \emline{18.50}{64}{1}{53.5}{4}{2}
 \put(10,35.0){\makebox(0,0)[lb]{$\sigma^\cZ_h$}}
 \put(37,5.0){\makebox(0,0)[lt]{$\sigma^\cZ_v$}}
 \put(55,4.5){\makebox(0,0)[lc]{$\sigma^\cA_v$}}
 \put(4.5,15.5){\makebox(0,0)[lt]{$\sigma^\cA_h$}}
 \put(46.5,41.5){\makebox(0,0)[cb]{0}}\put(46.5,26.5){\makebox(0,0)[ct]{1}}
 \put(37,47){\makebox(0,0)[lb]{2}}\put(56,47){\makebox(0,0)[rb]{3}}
\end{picture}
\caption{Nearest neighbors of the C-atom atom denoted by 0, are the
atoms 1, 2 and 3. Perpendicular to the figure at $\circ$ is $U$-axis,
while $\sigma^{\cZ/\cA}_{h/v}$ stands for vertical and horizontal
mirror planes of $\cZ$ and $\cA$ tubes.}\label{Fswcneigh}
\end{figure}

The tube $(n_1,n_2)$ is translationally periodic along the tube axis,
with period $a=a_0\sqrt{3q/2n\cR}$ ($a_0=2.461${\AA}), where $n$ is
the greatest common divisor of $n_1$ and $n_2$,
$q=2(n^2_1+n_1n_2+n^2_2)/n\mathcal{R}$ and $\cR=3$ if $(n_1-n_2)/n$
is multiple of 3 while $\cR=1$ otherwise. The tube is also invariant
under rotations for $2\pi/n$ around the tube axis. These
roto-translational symmetries form the group $\bL^{(1)}$. In
addition, chiral ($\cC$) tubes have horizontal $U$-axis (rotation for
$\pi$ around the $x$-axis; Fig. \ref{Fswcneigh}), doubling the former
group:
\begin{subequations}\label{Eline}\begin{equation}\label{ElineC}
{\bL}_{\cC}={\bT}^r_q{\bm D}_n=\bL^{(1)}+U\bL^{(1)}.
\end{equation}
Here, the parameter defining the screw-axis (helical) group
${\bT}^r_q$ is
$$r=\frac{q}{n}{\rm
Fr}\left[\frac{n}{q\mathcal{R}}(3-2\frac{n_1-n_2}{n_1})+\frac{n}{n_1}
\left(\frac{n_1-n_2}{n}\right)^{\varphi(\frac{n_1}{n})-1}\right]$$
(Fr denotes the fractional part and $\varphi$ the Euler function).
Zig-zag ($\cZ$, with $\cR=1$) and armchair ($\cA$, $\cR=3$) tubes,
$(n,0)$ and $(n,n)$ also have vertical mirror planes $\sigma_x$
(the other symmetry elements are generated from the mentioned
ones), which doubles the group $\bL_\cC$ (in these cases $q=2n$
and $r=1$):
\begin{equation}\label{ElineZA}
{\bL}_{\cZ\cA}={\bT}^1_{2n}{\bm D}_{nh}={\bL}_\cC+\sigma_x\bL_\cC.
\end{equation}\end{subequations}

The additional $U$ and $\sigma_x$ symmetries yield new conserved
quantum numbers: $+$ and $-$ ($A$ and $B$) denote even and odd
states with respect to the $U$-axis ($\sigma_x$ plane). The
electronic bands assignation by the complete set of conserved
quantum numbers makes easier theoretical predictions of their
degeneracies and selection rules relevant for the various physical
processes. Here we perform this task for the bands necessary to
discuss the conductivity.

The tight binding hamiltonian with single atomic $p$-orbital
$\ket{i}$ per $i$-th site encountering only the nearest neighbors
interaction is $H=\frac{1}{2}\sum_s\sum_{i=1}^3V_i\ket{s}\bra{s,i}$,
with $i$ running over the neighbors of the $s$-th C atom (Fig.
\ref{Fswcneigh}). Neglecting the local curvature, i.e. taking $V_i=V$
(estimated between $-3$eV and $-2.5$eV), the band structure of such
simplified hamiltonian $H^0$ can be found easily
\cite{HAM92,DRE98b,WHI93}. The assignation by the full line group
quantum numbers has been performed recently \cite{GP-7TBA}. In
particular, for the tubes with integer $(n_1-n_2)/3$, there are pairs
of bands crossing at Fermi level $E_\rF=0$. These bands and their
intersection point are characterized by the quantum numbers given in
table \ref{TQNF}.

\begin{table}[ht]\centering
\caption{The quantum numbers at Fermi level. Both cases of
$\tilde{m}_\rF$ are gathered in
$\tilde{m}_{\rF}\dot=\frac{2}{3}nr\cR$, with equality modulo the
interval $(-n/2,n/2]$.}\label{TQNF}

\begin{tabular}{ccccc|ccc} \hline
 &$\cC$:  $\cR=3$&$\cR=1$&$\cZ$&$\cA$&&$\cC$: $\cR=3$&$\cR=1$\\ \hline
 $k_{\rF}$&$\frac{2\pi}{3a}$&$0$&$0$&$\frac{2\pi}{3a}$&$\tilde{k}_F$&\multicolumn{2}{c}{$\frac{2q\pi}{3na}$}\\
 $m_{\rF}$&$nr\pymod{q}$&$\pm\frac{q}{3}$&$\frac{2n}{3}$&$n$&$\tilde{m}_{\rF}$&0&$\tfrac{n}{3}(-1)^{
 \mathrm{Fr}(r/3)}$\\ \hline
\end{tabular}\end{table}

For $\cC$ tubes, the simpler to handle with set of quantum numbers,
$\tk_\rF$ and $\tm_\rF$, is used. Introducing the angles
\begin{subequations}\label{EangY}\begin{eqnarray}
 &&\tilde\psi_1=-\tilde{k}a\frac{n_2}{q}+4r\pi\frac{2n_1+(1+r\cR)n_2}{3q},\\
 &&\tilde\psi_2= \tilde{k}a\frac{n_1}{q}+4r\pi\frac{(1-r\cR)n_1+2n_2}{3q},\\
 &&\tilde\psi_3=\tilde\psi_2-\tilde\psi_1,
\end{eqnarray}\end{subequations}
the Fermi bands for $\cC$, $\cZ$ and $\cA$ conducting tubes are:
\begin{subequations}\label{Eband}\begin{eqnarray}
&&\label{EbandC}E_\cC(\tk)=\pm V\sqrt{3+2\sum_i\cos\tilde\psi_i},\\
&&\label{EbandZ}G_\cZ(k)=\pm V\sqrt{2(1- \cos\frac{ka}{2})},\\
&&\label{EbandA}E^{A/B}_\cA(k)=\pm V(1-2\cos\frac{ka}{2}).
\end{eqnarray}\end{subequations}%
In the irreducible domain $\tk\in[0,q\pi/na]$ of the Brillouine zone
\cite{ALTMAN}, the both bands $E_\cC(\tk)$ correspond  to the same
two dimensional irreducible representation ${_{\tk}E}_{\tm_\rF}$ of
the group (\ref{ElineC}), \cite{IY-93}. On the contrary, two bands
$E^{A}_\cA(k)$ and $E^{A}_\cA(k)$ (in the irreducible domain
$k\in[0,\pi/a]$) correspond to different two dimensional
representations ${_kE}^A_n$ and ${_kE}^B_n$ of the group
(\ref{ElineZA}), \cite{GP-7TBA}. As for the $\cZ$ tubes, the both
bands $G_\cZ(k)$ correspond to the four fold representation
${_kG}_{2n/3}$ of the group (\ref{ElineZA}). These two bands cross at
$k_\rF=0$, giving raise to four fold degeneracy of Fermi level being
spanned by two pairs of states carrying representations
${_0E}^+_{2n/3}$ and ${_0E}^-_{2n/3}$ of different $U$ parities.

The non-crossing rule \cite{LANDAUIII}  asserts that the bands
assigned by the same values of complete set of symmetry based quantum
numbers may not cross (or touch). In fact, it can be easily derived
from the Wigner-Eckart theorem that in such a case more accurate
approach (retaining geometrical symmetry of the model) results in the
gap opening at the crossing points. Therefore, in a higher order
approximation, $\cC$ tubes attain a gap at $\tk_\rF$; the same
argument prevents crossing in any other $\tk$. Further, although at
$k_\rF=0$ the non-crossing rule at first glance does not obstruct the
conductivity of $\cZ$ tubes, the finite gap between the bands
$G^\pm_{\cZ}$ along the zone implies, by continuity argument, that a
gap arises at $k_\rF=0$, too. On the contrary, the relevant $\cA$
tube bands have different parities along Brillouine zone, being thus
beyond the scope of the non-crossing rule. In fact, in all these
cases the "accidental" degeneracy at $k_\rF$ reflects the increased
symmetry tacitly imposed by modeling the interatomic interactions by
mutually equal constants $V_i$: the hamiltonian became additionally
invariant under the permutations of the neighboring atoms. As the
following analysis shows, this artefact of the model is to disappear
in a more precise approach obeying the original symmetry.

The high symmetry of the nanotubes enables to get the exact
dispersion relations for the tight-binding calculations
\cite{GP-7TBA} with arbitrary chosen interaction coefficients $V_i$:
\begin{subequations}\label{Eband2}\begin{eqnarray}
&&\label{EbandC2}E_\cC(\tk)=\pm
\sqrt{\sum_i(V^2_i+2\frac{V_1V_2V_3}{V_i}\cos\tilde\psi_i)},\\
&&\label{EbandZ2}G_\cZ(k)=\pm \sqrt{V^2_1+V^2_2-
2V_1V_2\cos\frac{ka}{2}},\\
&&\label{EbandA2}E^{A/B}_\cA(k)=\pm (V_3-2V_1\cos\frac{ka}{2}).
\end{eqnarray}\end{subequations}%

\begin{figure}\centering
\includegraphics[width=12cm,height=3cm]{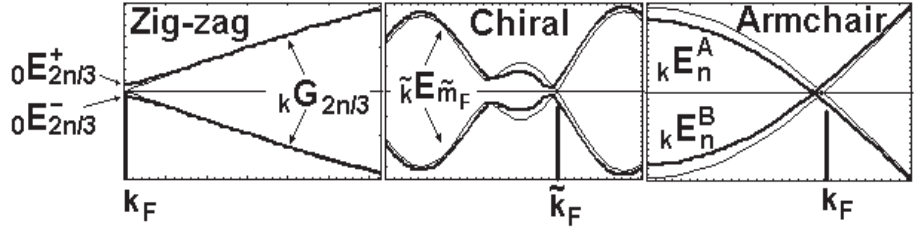}
\caption{\label{Fswcg2}Bands (\ref{Eband}) are crossed at $k_\rF$
(thin lines), while the corresponding bands (\ref{Eband2}) attain a
gap (thick lines), unless in the armchair case when the crossing
point is shifted to the left. The representations corresponding to
the bands are given.}
\end{figure}

To make the model realistic, the coefficients should be proportional
to the overlap integrals \cite{HARISON,JAPS} of the site $p^\bot$
orbitals, meaning that the first order correction to $V_i$ is
proportional to the bond length $d_i$ variation. Especially,
$V_0=V_1<V_2=V_3$ for $\cZ$ and $V_1=V_2<V_3$ for $\cA$ tubes. The
calculation made with such class of interaction confirm the
appearance of the gap (fig. \ref{Fswcg2}). It decreases with tube
diameter $D$ as $\beta/D^2$, with the coefficient $\beta$ depending
on the chiral angle. The armchair tubes preserve conductivity, since
their bands $E^A_\cA(k)$ and $E^B_\cA(k)$ cross in the new point
$k_\rF=(2/a) \arccos(V_3/2V_1)<2\pi/3a$.

To conclude, only for the armchair tubes the bands yielding the
conductivity within the simplest tight binding model have different
set of the conserved quantum numbers. The difference, the opposite
vertical mirror parity, can be observed only with help of the full
symmetry line group. In the view of the non-crossing rule,
exclusively armchair tubes conductivity is preserved, as illustrated
by more exact model. Regularity in the width of the opened gap in
$\cC$ and $\cZ$ tubes, as well as the shift of $k_\rF$ in $\cA$
tubes, can be used to resolve some ambiguities in the tubes
identification. It should be emphasized that the established complete
set of bands quantum numbers enables to calculate exhaustive set of
selection rules for the various processes involving interband
transitions. In this context it is relevant that the point $k_\rF$
given in the table \ref{TQNF} becomes weak van Hove singularity for
$\cC$ and $\cZ$ tubes in the presented model. Concerning $\cA$ tubes,
the new crossing point carries nontrivial helical $\tk$ and total
angular momentum $n\hbar$, although the "pure" angular momentum
vanishes; this property of the conducting band profoundly confirms
proposed chirality of the currents \cite{CCURENTS}.

\bibliographystyle{bibnotes}
\end{document}